\documentclass[conference,a4paper]{IEEEtran}

\usepackage{cite}

\ifCLASSINFOpdf{}
\usepackage[pdftex]{graphicx}
\else 
\usepackage[dvipdfmx]{graphicx}
\fi 
\usepackage{amsmath}
\usepackage{url}

\usepackage{bm}
\usepackage{amsfonts}
\usepackage{color}

\begin{document}
%
\title{Finding Association Rules by Direct Estimation of Likelihood Ratios}

\author{
  \IEEEauthorblockN{
    Kento Kawakami\IEEEauthorrefmark{1},
    Masato Kikuchi\IEEEauthorrefmark{1},
    Mitsuo Yoshida\IEEEauthorrefmark{1},
    Eiko Yamamoto\IEEEauthorrefmark{2},
    Kyoji Umemura\IEEEauthorrefmark{1}
  }
  \IEEEauthorblockA{
    \IEEEauthorrefmark{1}Department of Computer Science and Engineering \\
    Toyohashi University of Technology,
    Toyohashi 441--8580, Japan\\ \{k131820@edu,m143313@edu,yoshida@cs\}.tut.ac.jp, umemura@tut.jp
    \IEEEauthorblockA{
      \IEEEauthorrefmark{2}Department of Economics and Information \\
      Gifu Shotoku Gakuen University,
      Gifu 500--8288, Japan\\ eiko@gifu.shotoku.ac.jp
    }
  }
}


%

\IEEEoverridecommandlockouts{}
\IEEEpubid{\makebox[\columnwidth]{978--1--5386--3001--3/17/\$31.00~\copyright~2017 IEEE \hfill}
\hspace{\columnsep}\makebox[\columnwidth]{\hfill }}

\maketitle

\begin{abstract}
  In this paper, we propose a cost function that corresponds to the mean square errors between estimated values and true values of conditional probability in a discrete distribution.
  We then obtain the values that minimize the cost function.
  This minimization approach can be regarded as the direct estimation of likelihood ratios because the estimation of conditional probability can be regarded as the estimation of likelihood ratio by the definition of conditional probability.
  When we use the estimated value as the strength of association rules for data mining, we find that it outperforms a well-used method called Apriori.
\end{abstract}


%
\IEEEpeerreviewmaketitle{}

\section{Introduction}
Finding association rules between items is a fundamental operation in data mining~\cite{Ye:04,Datamining:11}.
Apriori~\cite{Agrawal:94} is an effective and efficient method for this operation.
This method uses the maximum likelihood estimator (MLE) of conditional probability $P(x \in R_c \mid y \in R_c)$ where $x$ and $y$ are variables for items, and $R_c$ is a random variable corresponding to a drawn record in a sampling space.
The event $x \in R_c$ indicates that drawn record $r_c$ (i.e., the sample of $R_c$) contains item $x$. 
Let $C(x,y)$ be the number of records that contain both item $x$ and item $y$.
This indicates that $C(x,y)$ is the frequency of co-occurrence of item $x$ and item $y$.
Let $C(x)$ be equivalent to $C(x,x)$.
The MLE of $P(x \in R_c \mid y \in R_c)$ becomes $C(x,y)/C(y)$.
It is well known that the value of MLE is unstable and may contain large amount of errors when $C(x,y)$ is small.
Apriori notices this problem and considers the strength of rules as zero when $C(x,y)$ is smaller than a certain threshold.
This threshold, called minimum support, is a type of parameter in Apriori, which should be tuned according to the data.

Although ignoring infrequently co-occurring items would be a natural, simple, and effective option for Apriori to find the strong association rules, the true strength of association for these items is actually not zero even when they co-occur in only a few records.
In addition, changing the treatment of the co-occurred item pairs on the boundary of the threshold may not be a fair treatment for all item pairs.

For this reason, we propose to use a cost function of errors, and the cost function reflects the mean square errors between true values and estimated values of conditional probability $P(x \in R_c \mid y \in R_c)$.
Moreover, we propose to use a normalized term in this cost function to overcome the instability of the estimation for some item pairs.
We then decide the estimated value that minimizes the cost function and consider the value as the strength of association of item pairs.

This method uses the framework of the direct estimation of likelihood ratios~\cite{Kanamori:09}, which is typically used for a continuous distribution.
This method requires the selection of kernel functions whose linear combination is the estimated result.
This selection is application dependent, and our selection for association rule mining is unique because our distribution is discrete.

We compared the effectiveness of the proposed method with that of Apriori, and found that the proposed method outperformed Apriori with statistical significance.
In addition to Apriori, we compared the proposed estimation with a smoothed estimator of probability.
This is because our analytical solution of cost minimization slightly resembles the formula of additive smoothing~\cite{Manning:99}, which contains Laplace smoothing in a special case.
We found that the smoothed estimator is worse than Apriori.
This suggests that the proposed method is novel and does not belong to this smoothed estimator.

\section{Related Work}
Apriori~\cite{Agrawal:94} is a standard method for finding the association rules. Apriori uses MLE with a threshold called ``minimum support'', as explained in the introduction.
Although the value of Apriori is famous for its efficient implementation that utilizes the threshold, we first focus our attention toward the effectiveness rather than efficiency.
Apriori achieves the effectiveness by ignoring the rules that may not exhibit sufficient evidence.

Predictive Apriori~\cite{Scheffer:05} obtains its effectiveness by using the expected value instead of MLE.
In this framework, the current database is considered as the sample.
First, Predictive Apriori learns the prior distribution of $C (x,y)/ C(y)$ over the choice of samples.
This prior distribution should be of the same shape for all $x$ and all $y$.
Predictive Apriori then estimates the posterior distribution of ratio $C (x,y)/C(y)$.
The posterior distribution should be affected by the $C (x,y)$ and $C (y)$.
Predictive Apriori computes the expected value of the ratio over the choice of samples in the posterior distribution.
As it is hard or unreasonable to assume some parametric distribution for $C (x,y)/C (y)$, a histogram is suggested for expressing the prior distribution.
As a histogram consists of many numbers in its shape, Predictive Apriori needs many parameters.
We may interpret that Predictive Apriori obtains its effectiveness by tuning many parameters.

The direct estimation of likelihood ratios~\cite{Kanamori:09} is reported to be more accurate than obtaining two estimations of probability and then utilizing its ratio.
It forms the cost function of estimation error of ratio rather than forming two cost functions for each probability.
Then, the cost function is empirically expressed by the observed data.
By assuming that the estimation of ratio is expressed by the linear combination of functions (called kernel functions), the solution of cost minimization problem provides the weights on each kernel function and the shape of the likelihood ratio.
The primary concern here is how to decide the set of kernel functions.
The suggested functions~\cite{Sugiyama:10,Sugiyama:12} are a family of Gaussian functions, which is only significant for a continuous distribution.

There is another approach for this task. Kikuchi et al.~\cite{kikuchi:16} argue that the cost where false association rules are selected is considerably higher than the cost where true association rules are not selected.
Rather than estimating the true value, they proposed to form a confidence interval of estimation of the strength, and utilize the lower boundary of the interval in practice.
Similar to Predictive Apriori, they assumed a prior distribution of estimator.
Unlike Predictive Apriori, they reported that a simple uniform distribution could be used as its prior distribution.
Instead they introduced another threshold for generating the interval of posterior distribution.

\section{Formalizing the Problem}
Suppose there is a database, whose record is a set of finite items.
Our sampling space is records including future data.
The number of such records is infinite.
Our sample is the record in the database at present.
The number of this record is finite.
Items are defined to be the labels in a record.
We will formulate a record as a set of items, and an item as a symbol.
Let $S$ be all the possible symbols or items in the database.
Let $R_c$ be the random variable for a record in the sampling space, and let $x$ and $y$ be the specified items.
Our objective is to estimate $P(x \in{} R_c \mid y \in{} R_c)$ from the sample and consider the estimated value as the strength of association rule $x \leftarrow y$.

\section{Proposed Method}

In this paper, we use $f(x) := g(x)$ to define $f$. To explain the proposed method, we define the following functions.
\begin{IEEEeqnarray}{rl}
  p (x,y):=&P (x \in{} R_c, y \in{} R_c),\\
  p (y):=&P (y \in{} R_c),\\
  r (x,y):=&P (x \in{} R_c | y \in{} R_c).
\end{IEEEeqnarray}
By the definition of conditional probability, we have the following formula as~\cite{Sugiyama:10}.
\begin{equation}
  r (x,y)=\frac{p(x,y)}{p(y)}.
\end{equation}
Then, we introduce a model of $\hat{r}_{\bm{\alpha}} (x,y)$ for the estimation of $r (x,y)$.
This is a linear combination of kernel functions $\phi_{ij} (x,y)$, where the variables $i$ and $j$ move in $S$ starting from 1 to $v (= \mid S \mid)$, where $\alpha_{ij}$ is some non-negative real number and $\bm{\alpha}$ is a vector whose $(i \times v + j)$-th element is $\alpha_{ij}$ and $\bm{\phi} (x,y)$ is a vector whose $(i \times v + j)$-th element is $\phi_{ij} (x,y)$.
\begin{IEEEeqnarray}{rl}
  \hat{r}_{\bm{\alpha}} (x,y):=&\bm{\alpha}^{\mathrm{T}} \bm{\phi} (x,y) \\
  =& \sum^{v}_{i=1}\sum^{v}_{j=1} \alpha_{ij} \phi_{ij} (x,y).
\end{IEEEeqnarray}
These kernel functions are different from~\cite{Sugiyama:10}.
We select the kernel function that can provide an independent value for each item.
Item $w_i$ is the $i$-th element of $S$, and item $w_j$ is the $j$-th element of $S$.
\begin{equation}
  \phi_{ij} (x,y):=\begin{cases}
      1~&~\left( x=w_i,y=w_j \right), \\
      0~&~otherwise.
    \end{cases}
\end{equation}
By the definition of $\phi_{ij} (x,y)$, we have the following formula.
\begin{IEEEeqnarray}{rl}
  \hat{r}_{\bm{\alpha}} (w_i,w_j)=&\sum^{v}_{i'=1} \sum^{v}_{j'=1} \alpha_{i'j'} \phi_{i'j'} (w_i,w_j) \\
  =&\alpha_{ij}.
\end{IEEEeqnarray}
Now, we can define the cost function $J_0(\bm{\alpha})$, which corresponds to some of the square errors for all pairs of items.
\begin{IEEEeqnarray}{rl}
  J_0 (\bm{\alpha}):=&\frac{1}{2} \sum^{v}_{i=1} \sum^{v}_{j=1} (\hat{r}_{\bm{\alpha}} (w_i,w_j)-r (w_i,w_j))^2 p (w_j) \\
  =&\frac{1}{2} \sum^{v}_{i=1} \sum^{v}_{j=1} \hat{r}_{\bm{\alpha}} (w_i,w_j)^2 p (w_j) \nonumber\\
  &-\sum^{v}_{i=1} \sum^{v}_{j=1} \hat{r}_{\bm{\alpha}} (w_i,w_j) p (w_i,w_j) + C,
\end{IEEEeqnarray}
\noindent where
\begin{equation}
  C:=\frac{1}{2} {\sum^{v}_{i=1} \sum^{v}_{j=1} r (w_i,w_j)p (w_i,w_j)}.
\end{equation}
As $C$ is a constant to $\bm{\alpha}$, the minimization of $J (\bm{\alpha})$ provides an identical solution with $J_0 (\bm{\alpha})$.
\begin{IEEEeqnarray}{rl}
  J (\bm{\alpha}):=&J_0 (\bm{\alpha}) - C \\
  =&\frac{1}{2} \sum^{v}_{i=1} \sum^{v}_{j=1} \hat{r}_{\bm{\alpha}} (w_i,w_j)^2 p (w_j) \nonumber\\
  &-\sum^{v}_{i=1} \sum^{v}_{j=1} \hat{r}_{\bm{\alpha}} (w_i,w_j) p (w_i,w_j).
\end{IEEEeqnarray}

From the observations (or current database), we can estimate $\hat{J} (\bm{\alpha})$ as follows.
\begin{IEEEeqnarray}{rl}
  \hat{J} (\bm{\alpha}):=&\frac{1}{2} \sum^{v}_{i=1} \sum^{v}_{j=1} \hat{r}_{\bm{\alpha}} (w_i,w_j)^2 \frac{C (w_j)}{N} \nonumber\\
  &-\sum^{v}_{i=1} \sum^{v}_{j=1} \hat{r}_{\bm{\alpha}} (w_i,w_j) \frac{C (w_i,w_j)}{N},
\end{IEEEeqnarray}
where $N$ is the number of records in the observation, $C (w_j)$ is the number of records that contain $w_j$, and $C (w_i,w_j)$ is the number of records that contain both $w_i$ and $w_j$.

Our optimization problem is to minimize $\hat{J} (\bm{\alpha})$ with the L2 normalization term.
In the following formula, $\lambda$ should be a positive number.
Considering that $J (\bm{\alpha})$ becomes small according to $1/N$, we consider the coefficients as $\lambda / N$, and the value of $\lambda$ will be a constant.
The reason for introducing the normalization term is to avoid overfitting to the observed data.
The problem is expressed as follows.

\begin{IEEEeqnarray}{c}
  \min_{\bm{\alpha \in{} \mathbb{R}^{v \times{} v}}} \left[ \hat{J} (\bm{\alpha}) + \frac{\lambda}{2N}\bm{\alpha}^{\mathrm{T}}\bm{\alpha} \right] \\
  {\rm subject~to~} \forall{} i, \forall{} j~\alpha_{ij} \geq{} 0. \nonumber
\end{IEEEeqnarray}

Fortunately, this problem exhibits an analytical solution because the solution with constraints is the same as the solution without constraints.

To obtain the solution without constraints, we use the following formula.
\begin{IEEEeqnarray}{l}
  \frac{\partial}{\partial \alpha_{ij}} \left(\hat{J} (\bm{\alpha}) + \frac{\lambda}{2N}\bm{\alpha}^{\mathrm{T}}\bm{\alpha} \right) = 0.
\end{IEEEeqnarray}
It provides the solution without constraints.
\begin{IEEEeqnarray}{rl}
  \alpha_{ij}&\left( \frac{C (w_j)}{N} + \frac{\lambda}{N} \right) -  \frac{C (w_i,w_j)}{N} = 0,\\
  \alpha_{ij}&=\frac{\frac{C (w_i,w_j)}{N}}{\frac{C (w_j)}{N} + \frac{\lambda}{N}},\\
  \alpha_{ij}&=\frac{C (w_i,w_j)}{C (w_j) + \lambda}.
\end{IEEEeqnarray}

This solution satisfies the constraints. Therefore, this is the solution with constraints.
In summary, we estimate the strength by using the following formula.
The value of $\lambda$ should be tuned using the learning dataset.
It should also be noted that if $\lambda = 0$ then the system becomes MLE.

\[
  \hat{r}_{proposed} (x,y) = C (x,y) / (C (y) + \lambda).
\]

\section{Experimental Setting and Evaluation Method}

For the experiment and evaluation, we selected $S$ as a set of city and prefecture names.
Each record of database corresponds to the symbols in $S$ in a newspaper article\footnote{\url{http://www.nichigai.co.jp/sales/mainichi/mainichi-data.html} (accessed 2017--6--15)} of seven years (1991 -- 1997).
We constructed 14 databases that correspond to newspaper articles of each half year, and we labeled each database by its year and an additional character.
TABLE~\ref{tbl:properties_of_datasets} shows each period, number of the articles, number of the true rules, and number of the all of the pairs in each dataset.
We assume that each newspaper article includes the description of several places of interest; thus, the name of the city or prefecture appears in the article with some probability.
It should also be noted that many labels of personal names could also be the labels of places.
This makes the finding association rules difficult or interesting.
Moreover, as we can judge whether the observed city is in the observed prefecture, it is possible to judge whether the mined association rule is correct or not.
As we have consensus for the relationship, the judgement agrees among all people.

\begin{table}[!t]
  \centering
  \caption{Properties of the Datasets\label{tbl:properties_of_datasets}}
  \begin{tabular}{l|l|r|r|r}
    \cline{2-5}
    & \multicolumn{1}{c|}{each period} & \multicolumn{1}{c|}{articles} & \multicolumn{1}{c|}{true rules} & \multicolumn{1}{c}{all of the pairs} \\ \hline
    \multicolumn{1}{l|}{91(a)} & 1/1/1991--~6/30/1991      &  25,510                  &     3,398                     &     187,994   \\ \hline
    \multicolumn{1}{l|}{91(b)} & 7/1/1991--12/31/1991     &  26,722                  &     2,969                     &     108,304   \\ \hline
    \multicolumn{1}{l|}{92(a)} & 1/1/1992--~6/30/1992      &  26,415                  &     2,827                     &     59,884    \\ \hline
    \multicolumn{1}{l|}{92(b)} & 7/1/1992--12/31/1992     &  30,172                  &     3,294                     &     202,157   \\ \hline
    \multicolumn{1}{l|}{93(a)} & 1/1/1993--~6/30/1993      &  25,704                  &     2,979                     &     86,700    \\ \hline
    \multicolumn{1}{l|}{93(b)} & 7/1/1993--12/31/1993     &  26,327                  &     3,251                     &     177,970   \\ \hline
    \multicolumn{1}{l|}{94(a)} & 1/1/1994--~6/30/1994      &  32,530                  &     3,183                     &     106,660   \\ \hline
    \multicolumn{1}{l|}{94(b)} & 7/1/1994--12/31/1994     &  33,392                  &     3,685                     &     214,703   \\ \hline
    \multicolumn{1}{l|}{95(a)} & 1/1/1995--~6/30/1995      &  37,856                  &     3,256                     &     148,010   \\ \hline
    \multicolumn{1}{l|}{95(b)} & 7/1/1995--12/31/1995     &  38,707                  &     3,206                     &     148,201   \\ \hline
    \multicolumn{1}{l|}{96(a)} & 1/1/1996--~6/30/1996      &  37,683                  &     3,218                     &     119,299   \\ \hline
    \multicolumn{1}{l|}{96(b)} & 7/1/1996--12/31/1996     &  20,854                  &     2,721                     &     88,143    \\ \hline
    \multicolumn{1}{l|}{97(a)} & 1/1/1997--~6/30/1997      &  42,668                  &     3,117                     &     112,795   \\ \hline
    \multicolumn{1}{l|}{97(b)} & 7/1/1997--12/31/1997     &  29,298                  &     2,699                     &     96,464                    
  \end{tabular}
\end{table}

As we do not have the correct answer for the relations among cities or among prefectures, we focus on the relation between a city and a prefecture.
In other words, we compute $\hat{r} (x,y)$, in the case where $x$ is the city name, and $y$ is the prefecture name.
The output of the system is the ranked list $(x, y)$ by $\hat{r} (x,y)$.
We eliminated the result in other cases, because we cannot determine whether the pair is correct or not.
By sorting according to the estimated value, the system outputs the ranked list of relations.
At each rank of the output, we compute the recall rate by a given value of rank.
This metrics is also reference~\cite{kikuchi:16}.

\[
  \rm{Recall = \frac{Number~of~true~positive}{Number~of~true~rules}}.
\]

As there are no systems that can detect the relationship for an unseen city and an unseen prefecture, the total number of true rules is the number of distinct cities times the number of distinct prefecture in the database used.
By plotting the recall rate for each rank, we can obtain the graph in Fig.~\ref{fig:exa_plot}, which will be explained later.
Moreover, we can compare the value of precision using the same graph. As precision is proportional to the slope toward the origination point, considering the number of total output is the definition of rank.
This metrics is also reference~\cite{kikuchi:16}.

\[
  \rm{Precision = \frac{Number~of~true~positive}{Number~of~total~output}}.
\]

When the plots line of two systems cross each other, we compare the system by the recall rate at a certain rank, for example, 1000, 4000 or 12000, where we can observe the difference among the systems in our experiment.

We compared three systems: Apriori, additive smoothing, and the proposed method.
Each system contains one parameter to tune.
In this experiment, as we have several datasets, the parameter is decided to be tuned in a certain year. Moreover, the value is used for obtaining the recall rate of the database of another year.

The estimator used by Apriori~\cite{Agrawal:94} is as follows.
In this formula, θ is the parameter of Apriori.
\[
  \hat{r}_{apriori} (x,y) = \begin{cases}
      C (x,y) / C (y) ~&~ ($C (x,y)$ \verb|>| \theta), \\
      0 ~&~ otherwise.
    \end{cases}
\]

The estimator used by additive smoothing~\cite{Manning:99} is as follows.
In this formula, $\mu$ is the parameter and $B$ is number of classes in the classification problem.
Our experiment contains two classes: one is the case where the pair is a true rule and the other is the case where the pair is a false rule.
Therefore, we set 2 to $B$. In theory, additive smoothing assumes that the prior distribution is the uniform distribution.
The parameter $\mu$ corresponds to the confidence in the prior distribution.
The value is the expected value of the posterior distribution.
Thus, if $\mu = 0$, the system becomes the MLE estimator.
If $\mu = 1$, the system becomes Laplace smoothing~\cite{Manning:99}.
\[
  \hat{r}_{additive\_smoothing} (x,y) = (C (x,y) + \mu) / (C (y) + \mu B).
\]

\subsection{Parameter Learning}
We tested three systems: Apriori, additive smoothing, and the proposed method, that is, the direct estimator of the likelihood ratio.
Let us call the proposed method ``Direct''.
As every system contains the same number of parameters, it is fair to compare with the same size of dataset.

We use the recall rate at rank 4000 for tuning the parameter.
We call this condition as TOP-4000.
Please note that this rate is also proportional to the precision rate at rank 4000 because both rates are proportional to the number of true positives, and the denominators of both ratios are constants at a fixed rank.
The number 4000 is selected for the condition that the results reveal the modest but the largest recall rate.

TABLE~\ref{tbl:search_variable} shows the values of the parameter for each system and each dataset.
We noticed that the dataset ``95(a)'' shows different values from other datasets for all the three systems.
This indicates that the distribution of the true rules in ``95(a)'' is different from other datasets.
As it affects all the three systems, we require the care for the treatment of database ``95(a)''.

\begin{table}[!t]
  \centering
  \caption{Variable of Maximum TOP-4000 Recall\label{tbl:search_variable}}
  \begin{tabular}{l|rr|rr|rr}
    \cline{2-7}
    &  \multicolumn{2}{c|}{Proposed} & \multicolumn{2}{c|}{Apriori} & \multicolumn{2}{c}{Additive smoothing} \\ \hline
    \multicolumn{1}{c|}{datasets} & \multicolumn{1}{c}{$\lambda$}   & \multicolumn{1}{c|}{Recall}  & \multicolumn{1}{c}{$\theta$}   & \multicolumn{1}{c|}{Recall}   & \multicolumn{1}{c}{$\mu$}    & \multicolumn{1}{c}{Recall}    \\ \hline
    91(a) & 5.500  & 0.3873  & 2 & 0.3655  & 9.001  & 0.3090 \\
    91(b) & 4.973  & 0.4816  & 3 & 0.4392  & 9.251  & 0.4025 \\
    92(a) & 7.667  & 0.5320  & 2 & 0.4970  & 9.750  & 0.4436 \\
    92(b) & 3.834  & 0.4083  & 2 & 0.3749  & 8.334  & 0.3509 \\
    93(a) & 5.001  & 0.5347  & 1 & 0.4918  & 9.000  & 0.4270 \\
    93(b) & 3.501  & 0.4457  & 2 & 0.4085  & 9.001  & 0.3562 \\
    94(a) & 4.056  & 0.4763  & 1 & 0.4461  & 9.000  & 0.3827 \\
    94(b) & 2.000  & 0.3848  & 1 & 0.3677  & 9.501  & 0.3083 \\
    95(a) & 1.319  & 0.3888  & 1 & 0.3900  & 2.501  & 0.3799 \\
    95(b) & 5.251  & 0.4361  & 1 & 0.3955  & 9.001  & 0.3637 \\
    96(a) & 4.091  & 0.4521  & 1 & 0.4276  & 9.500  & 0.3599 \\
    96(b) & 5.167  & 0.4972  & 1 & 0.4594  & 6.000  & 0.3866 \\
    97(a) & 6.302  & 0.4501  & 2 & 0.4277  & 9.001  & 0.3808 \\
    97(b) & 5.191  & 0.4794  & 2 & 0.4483  & 9.500  & 0.3783
  \end{tabular}
\end{table}

It should be noted that the obtained parameter is not zero for all the cases.
This indicates that all the three systems outperform the simple MLE, because all the three systems become MLE when the parameter is equal to zero.

\subsection{Evaluation Analysis}
First, we computed the recall rates at TOP-1000 (small recall rate condition), TOP-4000 (modest recall rate condition), and TOP-12000 (recall rate-oriented condition), by learning the parameters from the dataset of the previous period.
TABLE~\ref{tbl:experimental_result} shows the result.
For each dataset, the underlined numbers show the best result system among proposed method, Apriori and additive smoothing.
Among the three systems, the proposed method shows the best result. 
Compared with Apriori using one sided paired t-test, the $\alpha$-level of significance is 0.0002 in the TOP-4000 condition, and 0.002 in the TOP-12000 condition.
Compared with additive smoothing, the α-level of significance is extremely small in both the TOP-4000 and TOP-12000 conditions.
In the TOP-1000 condition, the difference is not apparent because all the three systems behave similar to MLE, when $C (x,y)$ is large.

\begin{table*}[!t]
  \centering
  \caption{Experimental Result: Recall Rate at TOP-1000, TOP-4000 and TOP-12000\label{tbl:experimental_result}}
  \begin{tabular}{l|rrr|rrr|rrr}
    \cline{2-10}
    \multicolumn{1}{l|}{}   & \multicolumn{3}{c|}{TOP-1000}  & \multicolumn{3}{c|}{TOP-4000}         & \multicolumn{3}{c}{TOP-12000}       \\ \hline
    \multicolumn{1}{c|}{period's} & \multicolumn{1}{c|}{Proposed} & \multicolumn{1}{c|}{Apriori} & \multicolumn{1}{c|}{Additive} & \multicolumn{1}{c|}{Proposed} & \multicolumn{1}{c|}{Apriori} & \multicolumn{1}{c|}{Additive} & \multicolumn{1}{c|}{Proposed} & \multicolumn{1}{c|}{Apriori} & \multicolumn{1}{c}{Additive} \\ \hline
    91(b) & \underline{0.2385}  & 0.2082  & 0.2307  & \underline{0.4709}  & 0.4271  & 0.4022  & \underline{0.6831}  & 0.5935  & 0.5517 \\
    92(a) & 0.2681  & 0.2402  & \underline{0.2706}  & \underline{0.5179}  & 0.4436  & 0.4429  & \underline{0.7503}  & 0.4981  & 0.6625 \\
    92(b) & 0.2177  & 0.2077  & \underline{0.2219}  & \underline{0.3986}  & 0.3749  & 0.3509  & \underline{0.5698}  & 0.5088  & 0.4350 \\
    93(a) & \underline{0.2706}  & 0.2625  & 0.2645  & \underline{0.5183}  & 0.4703  & 0.4267  & \underline{0.7392}  & 0.5740  & 0.6284 \\
    93(b) & 0.2190  & 0.1529  & \underline{0.2218}  & \underline{0.4343}  & 0.4002  & 0.3559  & \underline{0.6026}  & 0.5752  & 0.4848 \\
    94(a) & 0.2114  & 0.2325  & \underline{0.2334}  & \underline{0.4753}  & 0.4461  & 0.3827  & \underline{0.6956}  & 0.5928  & 0.5680 \\
    94(b) & 0.1818  & 0.1194  & \underline{0.1837}  & \underline{0.3845}  & 0.3677  & 0.3069  & \underline{0.5666}  & 0.5341  & 0.4334 \\
    95(a) & 0.1904  & \underline{0.2187}  & 0.1729  & 0.3836  & \underline{0.3900}  & 0.3744  & \underline{0.5952}  & 0.5534  & 0.5344 \\
    95(b) & \underline{0.2280}  & 0.1850  & 0.2274  & \underline{0.4017}  & 0.3955  & 0.3584  & 0.5708  & \underline{0.5923}  & 0.4841 \\
    96(a) & 0.2057  & 0.1830  & \underline{0.2216}  & \underline{0.4422}  & 0.4276  & 0.3595  & \underline{0.6426}  & 0.6165  & 0.5466 \\
    96(b) & \underline{0.2407}  & 0.2117  & 0.2367  & \underline{0.4914}  & 0.4594  & 0.3863  & \underline{0.7144}  & 0.6516  & 0.5752 \\
    97(a) & 0.2060  & 0.1562  & \underline{0.2246}  & \underline{0.4485}  & 0.4103  & 0.3786  & \underline{0.6782}  & 0.6365  & 0.5496 \\
    97(b) & \underline{0.2386}  & 0.2156  & 0.2308  & \underline{0.4757}  & 0.4483  & 0.3779  & \underline{0.6810}  & 0.5891  & 0.5502\\
    \hline
    average & 0.2243 & 0.1995  & 0.2262 & 0.4495 & 0.4201 & 0.3772 & 0.6530 &0.5782 & 0.5388  \\
    standard deviation & 0.0268 & 0.0393 & 0.0264 &0.0472 & 0.0329 & 0.0345 &0.0658  &0.0455 & 0.0670  
  \end{tabular}
\end{table*}

\begin{figure}[!t]
  \centering
  \includegraphics[width=3.0in]{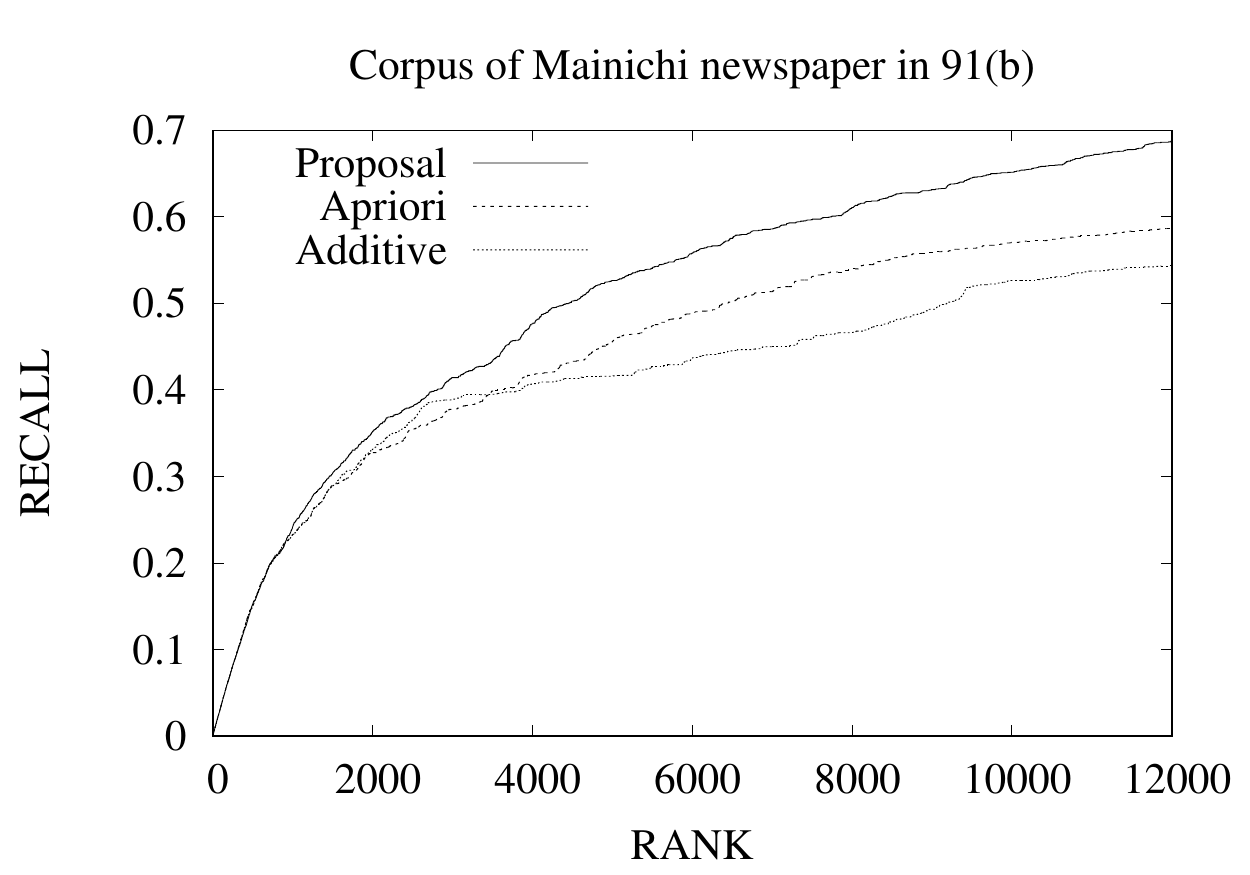}
  \caption{Plot of recall at rank, which is learned parameter in 91(a), and tested in 91(b). There is no significant difference in rank 0--2000. The proposed method has the highest recall rate in rank 2000--12000.\label{fig:exa_plot}}
\end{figure}

Fig.~\ref{fig:exa_plot} is the plot of recall rate at rank, which is a learned parameter in 91(a), and a tested parameter in 91(b).
As the curves do not cross each other, the proposed method is recommended to be used for all range of ranks.

\subsection{Discussion}
It is difficult to compare the proposed system with Predictive Apriori owing to the difference in the number of parameters.
If we need to learn the prior distribution in the shape of a histogram, we require at least ten or more parameters.
In general, a system that contains many parameters may show good performance but tuning may be difficult.

If we assume the beta distribution~\cite{Wasserman:04} as the prior distribution of the probability of the Bernoulli trial, the posterior distribution also becomes the beta distribution.
Then, the expected value of the posterior distribution is $(C (x,y) + \alpha) / (C (y) + \alpha + \beta)$ where $\alpha$ and $\beta$ are parameters in the beta distribution.
By assuming the prior distribution as beta distribution, Predictive Apriori becomes a system that is similar to the proposed system.

When $\alpha = 1$ and $\beta = 1$, the beta distribution becomes the uniform distribution of the [0,1] range.
The expected value of the corresponding posterior distribution provides the Laplace smoothing result.
If we introduce the strength of belief in this framework, we can obtain the additive smoothing results.
Therefore, we may consider additive smoothing as a parameter version of Predictive Apriori.

It should be noted that the beta distribution requires $\alpha \verb|>| 0$.
This indicates that our proposed method cannot be the special case of Predictive Apriori with the beta prior distribution.
We cannot imagine a reasonable prior distribution corresponding to our proposed estimation.

Kikuchi et al.~\cite{kikuchi:16} propose to use the confidence interval by assuming that the prior distribution is a uniform distribution.
They also contain one parameter to tune, which is the confidence level.
These intervals should be decided numerically for each confidence level, $C (x,y)$ and $C (y)$.
This indicates that tuning the confidence level is difficult because of the total amount of computation, and practical impossibilities. 
Using the provided table, Kikuchi's method shows almost the same result as our system.

\section{Conclusion}
In this paper, we proposed a cost function corresponding to the mean square errors between estimated values and true values of conditional probability in a discrete distribution.
We then obtained the values that minimized the cost function.
This minimization approach can be considered as a direct estimation of likelihood ratios.

We compared the proposed method with Apriori using 14 datasets.
By comparing the recall rates at TOP-4000, we can observe the modest but largest recall rate.
We measured the recall rates at TOP-1000, TOP-4000, and TOP-12000 by using the previous year database for learning the parameter of each system.
The average of recall rates at these conditions reveals that the proposed method outperforms Apriori.
We can observe the statistical significance by using the one-sided t-test.

The optimal parameter value for dataset ``95(a)'' requires more examination. This is because it may provide a hint to reveal the hidden problem of our framework, and this is our future investigation.
Moreover, we endeavor to use another method of smoothing for estimating this conditional probability in our future investigation.

\bibliography{ICAICTA2017}
\bibliographystyle{IEEEtran}



%

\end{document}